\documentclass[10pt]{article}
\usepackage{fancyhdr}
\usepackage{extramarks}
\usepackage{amsmath}
\usepackage{amsthm}
\usepackage{amsfonts}
\usepackage{siunitx}
\usepackage{tikz}
\usepackage[plain]{algorithm}
\usepackage{algpseudocode}
\usepackage{multirow}
\usepackage{booktabs}
\usepackage{palatino}
\usepackage{graphicx}
\usepackage{subfigure}
\usepackage[colorlinks,linkcolor=black,anchorcolor=black,citecolor=black,urlcolor=blue]{hyperref}
\usepackage{amsmath,bm}
\usepackage{booktabs}
\usepackage{mathtools}
\usepackage{amssymb}
\usepackage{caption}
\usepackage{capt-of}
\usepackage{mciteplus}
\usepackage{cite}
\usepackage{mathrsfs}
\usepackage[title,titletoc,toc]{appendix}
\usepackage{xr}
\usepackage{parskip}
\usepackage{soul}
\usepackage{textcomp}
\usepackage[colaction]{multicol}
\usepackage[switch]{lineno}
\usepackage{lipsum}
\usepackage{etoolbox}
\usepackage{longtable}
\usepackage{array}
\usepackage{tablefootnote}
\usepackage{ragged2e}
\newcolumntype{C}[1]{>{\centering\arraybackslash}p{#1}}
\usetikzlibrary{automata,positioning}
\usetikzlibrary{automata,positioning}

\begin{document}

\title{The evolution of the mechanisms of SARS-CoV-2 evolution revealing vaccine-resistant mutations in Europe and America} 
\author{Rui Wang$^1$, Jiahui Chen$^1$ and Guo-Wei Wei$^{1,2,3}$\footnote{
 		Corresponding author.		Email: weig@msu.edu} \\
 $^1$ Department of Mathematics, \\
 Michigan State University, MI 48824, USA.\\
 $^2$ Department of Electrical and Computer Engineering,\\
 Michigan State University, MI 48824, USA. \\
 $^3$ Department of Biochemistry and Molecular Biology,\\
 Michigan State University, MI 48824, USA. \\
 }
\date{\today} 

\maketitle

\begin{abstract}
The importance of understanding SARS-CoV-2 evolution cannot be overemphasized. Recent studies confirm that natural selection is the dominating mechanism of SARS-CoV-2 evolution, which favors mutations that strengthen viral infectivity. We demonstrate that vaccine-breakthrough or antibody-resistant mutations provide a new mechanism of viral evolution.  Specifically, vaccine-resistant mutation Y449S in the spike (S) protein receptor-bonding domain (RBD), which occurred in co-mutation [Y449S, N501Y], has reduced infectivity compared to the original SARS-CoV-2 but can disrupt existing antibodies that neutralize the virus. By tracing the evolutionary trajectories of vaccine-resistant mutations in over 1.9 million SARS-CoV-2 genomes, we reveal that the occurrence and frequency of vaccine-resistant mutations correlate strongly with the vaccination rates in Europe and America. 
We anticipate that as a complementary transmission pathway, vaccine-resistant mutations will become a dominating mechanism of SARS-CoV-2 evolution when most of the world's population is vaccinated. Our study sheds light on SARS-CoV-2 evolution and transmission and enables the design of the next-generation mutation-proof vaccines and antibody drugs.

\end{abstract}
 \newpage

\setcounter{page}{1}
\renewcommand{\thepage}{{\arabic{page}}}


\section{Introduction}
Started in late 2019, the coronavirus disease 2019 (COVID-19) pandemic caused by severe acute respiratory syndrome coronavirus 2 (SARS-CoV-2) has had devastating impacts worldwide, which has plunged the world into an economic recession. Although several authorized vaccines have offered promise to control the disease in early 2021, the emergence of multiple variants of SARS-CoV-2 indicates that the combat with SARS-CoV-2 will be protracted. At this stage, almost all SARS-CoV-2 vaccines and monoclonal antibodies (mAbs) are targeted at the spike (S) protein \cite{malik2021targets}, while mutations on the S protein have been verified to link to the efficacy of existing vaccines and viral infectivity \cite{annavajhala2021novel,chen2021prediction}. Therefore, it is imperative to understand the mechanisms of viral mutations, especially on the S gene of SARS-CoV-2, which will promote the development of mutation-proof vaccines and mAbs.

\begin{figure}[ht!]
	\centering
	\includegraphics[width = 0.95\textwidth]{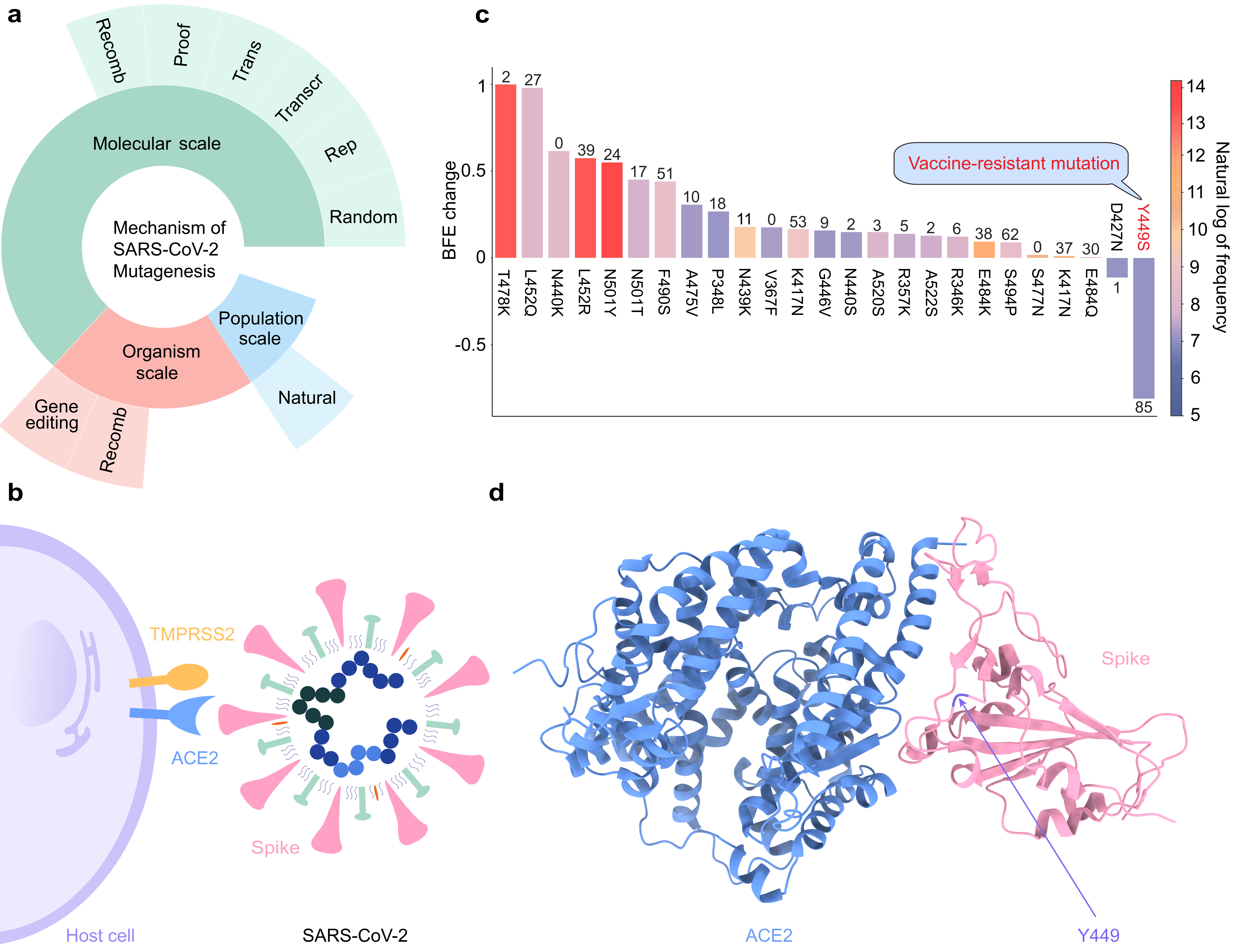}
	\caption{{\bf a} The mechanism of mutagenesis. Nine mechanisms are grouped into three scales: 1) molecular-based mechanism (green color); 2) organism-based mechanism (red color); 3) population-based mechanism (blue color). The random shifts (Random), replication error (Rep), Transcription error (Transcr), viral proofreading (Proof), and recombination (Recomb) are the six molecular-based mechanisms. The gene editing and the host-virus recombination are the organism-based mechanism. In addition, the natural selection (Natural) is the population-based mechanism, which is the mainly driven source in the transmission of SARS-CoV-2. {\bf b} A sketch of SARS-CoV-2 and its interaction with host cell. {\bf c}  Illustration of 25 single-site RBD mutations with top frequencies. The height of each bar shows the BFE change of each mutation, the color of each bar represents the natural log of frequency of each mutation, and the number at the top of each bar means the AI-predicted number of antibody and RBD complexes that may be significantly disrupted by a single site mutation.  {\bf d} Illustration of SARS-CoV-2 S protein with human ACE2. The blue chain represents the human ACE2, the pink chain represents the S protein, and the purple fragment on the S protein points out the two vaccine-resistant mutations Y449S/H.}
	\label{fig:mechanism}
\end{figure}

 The mechanism of mutagenesis is driven by various competitive processes \cite{sanjuan2016mechanisms,grubaugh2020making,kucukkal2015structural,yue2005loss,wang2020host}, which can be categorized into 3 different scales with many factors as illustrated in \autoref{fig:mechanism} {\bf a}: 1) the molecular scale, 2) the organism scale, and  3) the population scale. From the molecular-scale perspective, the random shifts, replication errors, transcription errors, translation errors, viral proofreading, and viral recombination are the main driven sources. Moreover, the host gene editing induced by the adaptive immune response \cite{wang2020host} and the recombination between the host and virus are the key-driven factors at the organism level. Furthermore, the natural selection popularized by Charles Darwin is a critical process, which favors mutations that have reproductive advantages for the virus to have adaptive traits in evolution. Such complicated mechanisms of viral mutagenesis make the comprehension of viral transmission and evolution a grand challenge.  

Although there are 28,780 unique single mutations distributed evenly on the whole SARS-CoV-2 genome, the mutations on the  S  gene stand out among all 29 genes on SARS-CoV-2 due to the mechanism of viral infection. Under assistant with host transmembrane protease, serine 2 (TMPRSS2), SARS-CoV-2 enters the host cell by interacting with its S protein and the host angiotensin-converting enzyme 2 (ACE2) \cite{hoffmann2020sars} (See \autoref{fig:mechanism} {\bf b}). Later on, antibodies will be generated by the host immune system, aiming to eliminate the invading virus through direct neutralization or non-neutralizing binding \cite{chen2020review,chen2021sars}, which makes the S protein the main target for the current vaccines. Specifically, there is a short immunogenic fragment located on the S protein of SARS-CoV-2 that can facilitate the SARS-CoV-2 S protein binding with ACE2, which is called the receptor-binding domain (RBD) \cite{tai2020characterization}. Studies have shown that the binding free energy (BFE) between the S RBD and the ACE2 is proportional to the infectivity \cite{li2005bats,qu2005identification,song2005cross,hoffmann2020sars,walls2020structure}. Therefore, tracking and monitoring the RBD mutations and their corresponding BFE changes will expedite understanding the infectivity, transmission, and evolution of SARS-CoV-2, especially for the new SARS-CoV-2 variants, such as Alpha, Beta, Gamma,  Delta, and Lambda, etc. \cite{yin2020genotyping}

 The current prevailing variants Alpha, Beta, Gamma, Delta, Kappa, Theta, Lambda, and Mu carry at least one vital mutation at residues 452 and 501 on the S RBD. Notably, in July 2020, we successfully predicted that residues 452 and 501 "have high changes to mutate into significantly more infectious COVID-19 strains" \cite{chen2020mutations}. In the same work, we hypothesized that ``natural selection favors those mutations that enhance the viral transmission" and provided the first evidence for infectivity-based natural selection. In other words, we revealed the mechanism of SARS-CoV-2 evolution and transmission based on very limited genome data in July 2020 \cite{chen2020mutations}. Additionally, we predicted three categories of RBD mutations: 1)  most likely (1149 mutations), 2) likely (1912 mutations), and 3) unlikely (625 mutations) \cite{chen2020mutations}. Up to now, all of the RBD mutations we detected fall into our first category \cite{chen2021prediction, wang2021vaccine}. Until now, all of the top 100 most observed RBD mutations have BFE change greater than the average BFE changes of -0.28kcal/mol (the average BFE changes for all RBD mutations\cite{wang2021emerging}). It is an extremely low odd (i.e., $\frac{1}{1.27\times 10^{30}}$) for 100 RBD mutations to accidentally have BFE changes simultaneously above the average value, which confirms our hypothesis that the transmission and evolution of new SARS-CoV-2 variants are governed by infectivity-based natural selection, despite all other competing mechanisms \cite{chen2020mutations}.  
 Our predictions rely on algebraic topology \cite{carlsson2009topology,edelsbrunner2000topological,xia2014persistent}-assisted deep learning \cite{wang2020topology,chen2020mutations}, but have been extensively validated \cite{chen2021prediction,chen2021revealing}.  
 However,   infectivity is not the only transmission pathway that governs viral evolution. Vaccine-resistant mutations or more precisely, antibody-resistant mutations,   that can disrupt the protection of antibodies has become a viable mechanism for new variants to transmit among the vaccinated population since the vaccine was put on the market.  In early January 2021, we have predicted that RBD mutations W353R, I401N, Y449D, Y449S, P491R, P491L, Q493P, etc.,  will weaken most antibody bindings to the S protein \cite{chen2021prediction}. Later on, we have provided a list of most likely vaccine escape RBD mutations with high frequency, including S494P, Q493L, K417N, F490S, F486L, R403K, E484K, L452R, K417T, F490L, E484Q, and A475S \cite{wang2021vaccine}. Moreover, we have pointed out that Y449S and Y449H are two vaccine-resistant mutations, and ``Y449S, S494P, K417N, F490S, L452R, E484K, K417T, E484Q, L452Q, and N501Y" are the top 10 mutations that will disrupt most antibodies with high-frequency \cite{wang2021emerging}. As mentioned in Ref. \cite{clark2021sars}, RBD mutations such as E484K/A, Y489H, Q493K, and N501Y found in late-stage evolved S variants ``confer resistance to a common class of SARS-CoV-2 neutralizing antibodies", which suggests the viral evolution is also regulated by vaccine-resistant mutations. 

The objective of this work is to analyze the evolution of the mechanisms of SARS-CoV-2 evolution, driven by complementary viral transmission pathways. 
We demonstrate how the interplay among molecular-scale, organism-scale, population-scale mechanisms of SARS-CoV-2 mutations have affected the evolution of SARS-CoV-2. As a primary driven source of mutagenesis, the molecular-based mechanisms such as random shifts, transcription errors, proofreading, etc., changing the genetic information initially. Next, gene editing takes charge of the organism-based mechanism, suggesting the host immune response to the virus \cite{wang2020host}. Then, the population-level mechanism governs the transmission pathways of viral evolution. Two complementary pathways (infectivity  and vaccine-resistance) regulated by natural selection become the preponderance of evolution-driven force. The RBD mutations regulated by infectivity-based pathways exist in the prevailing variants, while the mutations regulated by the vaccine-resistant pathway start to emerge in countries with relatively high vaccination rates. In this work, 1,983,328 complete SARS-CoV-2 genomes that isolate from patients are decoded by single nucleotide polymorphism (SNP) calling, from where a total of 28,780 unique single mutations are detected. Among them, 737 RBD mutations are discovered up to September 20, 2021 (The detailed information can be found in the Supporting Information S5). Based on our comprehensive topology-based artificial intelligence (AI) model to predict RBD mutation-induced BFE changes of  RBD and ACE2/antibody complexes \cite{chen2020mutations, chen2021prediction}, the transmission trajectory of vaccine-resistant RBD mutations will be analyzed (The detailed information about methods and model can be viewed in the Supporting Information S1 and S2). Moreover, vaccine-resistant RBD mutation Y449S that has been found in more than 1000 isolates will be discussed. Furthermore, the vaccination rates of 12 countries where Y449S is distributed are also analyzed, which provides a reliable  explanation of the relation between the emergence of vaccine-resistant mutations and the vaccination rate. Such understanding of two complementary transmission pathways will shed light on the long-term efficacy of S-targeted antibodies countermeasures and benefit the development of next-generation mutation-proof vaccines and mAbs.

\section{Results}

\subsection{Evolutionary trajectories of viral RBD single mutations}
Studying the mechanisms of SARS-CoV-2 mutagenesis is beneficial to the understanding of viral transmission and evolution. 
The mainly driven force of viral evolution is regulated by natural selection, which is employed by two complementary transmission pathways: 1) infectivity-based pathway and  2) vaccine-resistant pathway. We have discussed the infectivity-based pathways in Ref.\cite{wang2021emerging} and \cite{chen2021review}. This section focuses on the vaccine-resistant pathway and its impact on the transmission and evolution of SARS-CoV-2. To understand the mechanisms of vaccine-resistant mutations, we first analyze 1,983,328 complete SARS-CoV-2 genomes, and a total of 28,780 unique single mutations are decoded. Among them, there are 737 non-degenerate RBD mutations. The infectivity of SARS-CoV-2 is proportional to the BFE between the S RBD and ACE2 \cite{li2005bats,qu2005identification,song2005cross,hoffmann2020sars,walls2020structure}.  Therefore, the BFE change induced by a specific RBD mutation reveals whether the RBD mutation is an infectivity-strengthen or an infectivity-weaken mutation. Similarly, the BFE change between S RBD and antibody induced by a given mutation reveals whether this mutation will strengthen the binding between S and antibody or not. Up to now, we have collected 130 antibody structures (see the Supporting Information S4), which includes Food and Drug Administration (FDA)-approved mAbs from Eli Lilly and Regeneron. For a specific RBD mutation, its antibody disruption count shows the number of antibodies that have antibody-S BFE changes smaller than -0.3 kcal/mol. The ACE2-S and antibody-S BFE changes induced by RBD mutations are predicted from our TopNetTree model \cite{chen2020mutations}, which is available at \href{https://github.com/WeilabMSU/TopNetmAb}{TopNetmAb}. All of the predicted BFE changes induced by RBD mutations can be found at \href{https://weilab.math.msu.edu/MutationAnalyzer/}{Mutation Analyzer}. \autoref{fig:mechanism} {\bf c} illustrates the top 25 most observed RBD mutations.  The height and color of each bar represent the ACE2-S BFE changes and frequency of each RBD mutation. The number at the top of each bar shows the antibody disruption count of each mutation. The detailed information can be viewed in Supplementary Information S4. It can be seen that 23 mutations have positive ACE2-S BFE changes, suggesting they are regulated by the infectivity-based transmission pathway. Howbeit, 2 RBD mutations D427N and Y449S, have negative BFE changes. Notably, mutation Y449S has a significantly negative BFE change  (-0.8112 kcal/mol) and a pretty large antibody disruption count (89), revealing a non-typical mechanism of mutagenesis. Such a mutation with significantly negative ACE2-S BFE change together with a high antibody disruption count is called a vaccine-resistant or antibody-resistant mutation. \autoref{fig:mechanism} {\bf d} is the illustration of SARS-CoV-2 S protein (blue color) with human ACE2 (pink color), and the Y449 residue (purple color) is located on the random coil of the S protein. Among all of the vaccine-resistant mutations, Y449S has the highest frequency (1189). In addition, at residue 449, mutations Y449H, Y449N, Y449D are all vaccine-resistant mutations that have been observed in more than 20 SARS-CoV-2 genome isolates.

\begin{figure}[ht!]
	\centering
	\includegraphics[width = 0.8\textwidth]{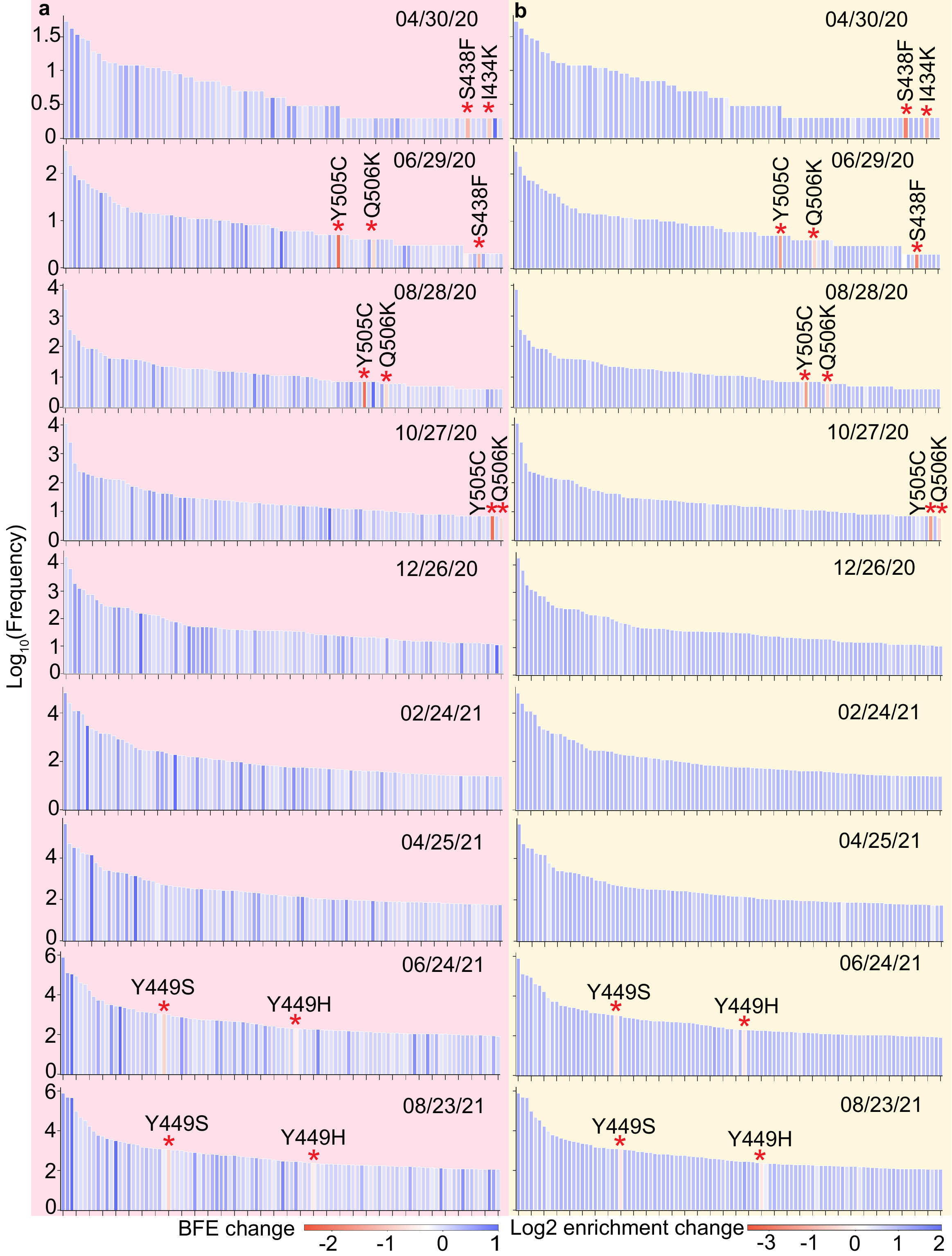}
	\caption{Most significant RBD mutations. 
	{\bf a} Time evolution of RBD mutations with its mutation-induced BFE changes per 60-day from April 30, 2020, to August 31, 2021. Here, only the top 100 most observed RBD mutations are displayed. The height and color of each bar represent the log frequency and ACE-S BFE change induced by a given RBD mutation. The red star marks the vaccine-resistant mutations with significantly negative   BFE changes. {\bf b} Time evolution of RBD mutations with its experimental mutation-induced log2 enrichment ratio changes per 60-day from April 30, 2020, to August 31, 2021. The height and color of each bar represent the log frequency and enrichment ratio change induced by a given RBD mutation. The red star marks vaccine-resistant mutations with significantly negative  BFE changes.
}
	\label{fig:barplot single}
\end{figure}

 To track the evolution trajectory of vaccine-resistant mutations, the BFE changes, log2 enrichment ratios \footnote{Log2 enrichment ratio is collected from the experimental deep mutation enrichment data in Ref. \cite{linsky2020novo}}, and log10 frequencies of RBD mutations are analyzed from April 30, 2020, to August 23, 2021, in every 60 days, as illustrated in \autoref{fig:barplot single}. Here, the top 100 most observed RBD mutations are displayed. In \autoref{fig:barplot single} {\bf a}, red stars mark the vaccine-resistant mutations that have negative BFE changes. Although a few vaccine-resistant mutations S438F, I434K, Y505C, and Q506K were detected before November 2020, they had relatively low frequencies. However, since December 2020, such vaccine-resistant mutations were no longer in the top 100 most observed RBD mutation list, suggesting that in this period, the evolution of SARS-CoV-2 is mainly regulated by natural selection through the infectivity-based transmission pathway. Notably, in May 2021, two vaccine-resistant mutations Y449S and Y449H, came back to the top 100 most observed RBD mutation list. In addition, Y449S has a relatively high frequency. Such finding indicates that natural selection not only favors those mutations that enhance the transmission but also those mutations that can disrupt plenty of antibodies since SARS-CoV-2 vaccines started to provide protection among populations in early May. Similarly, patterns can be found in \autoref{fig:barplot single} {\bf b}, suggesting our AI-predicted BFE changes are highly consistent with the deep mutational enrichment ratio from experiments \cite{linsky2020novo}.

\subsection{Evolutionary trajectories of viral RBD co-mutations}
\begin{figure}[ht!]
	\centering
	\includegraphics[width = 0.95\textwidth]{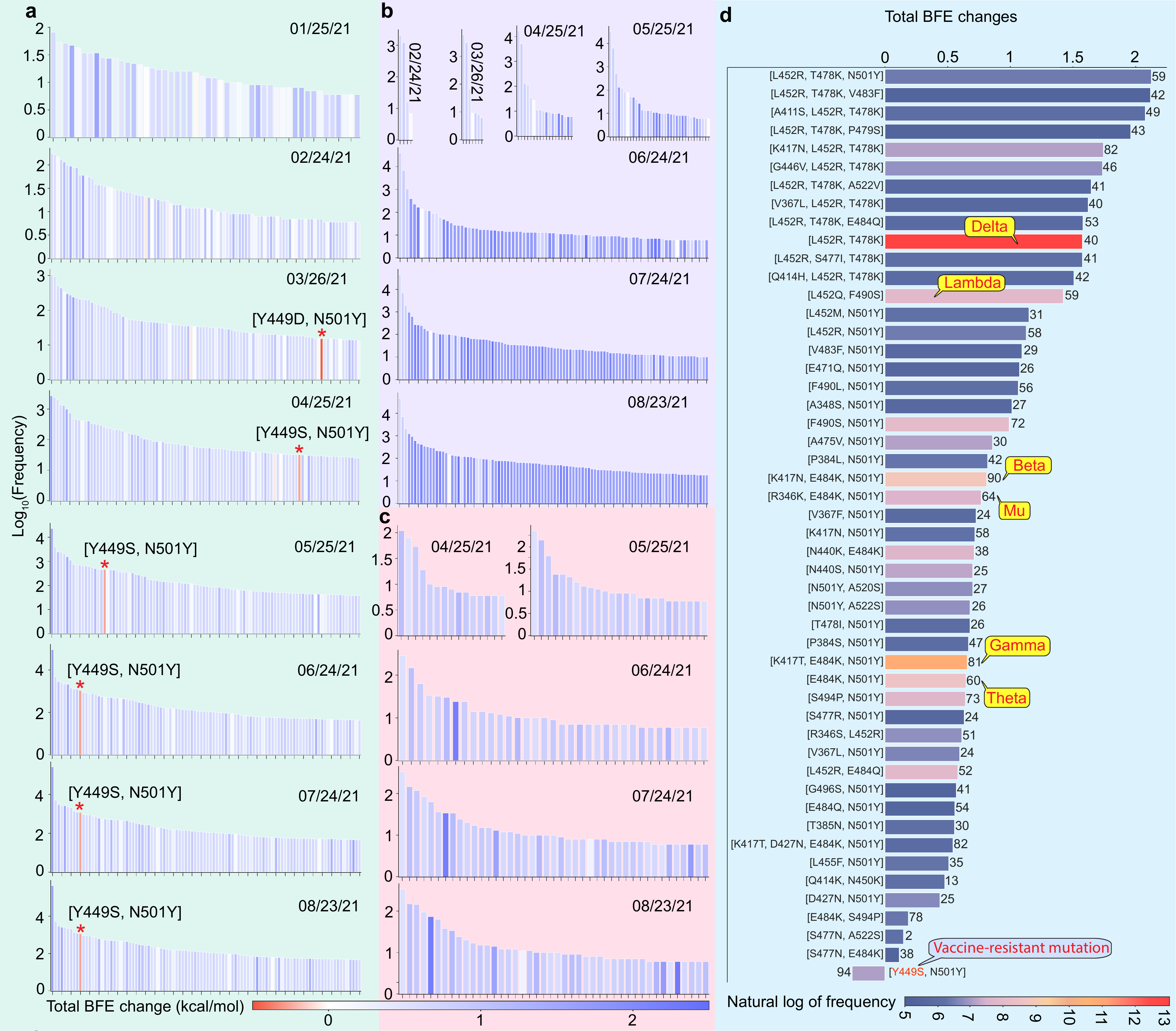}
	\caption{ RBD co-mutation analysis. {\bf a} Time evolutionary trajectory of RBD 2 co-mutations with its mutation-induced BFE changes per 30-day from January 25, 2021, to August 23, 2021. The height and color of each bar represent the log frequency and ACE-S BFE change induced by a given RBD mutation. Red stars mark the 2 co-mutations with significantly negative BFE changes. {\bf b} Time evolutionary trajectory of RBD 3 co-mutations with its mutation-induced BFE changes per 30-day from February 24, 2021, to August 23, 2021. The height and color of each bar represent the log frequency and ACE-S BFE change induced by a given RBD mutation.  {\bf c} Time evolutionary trajectory of RBD 4 co-mutations with its mutation-induced BFE changes per 30-day from April 25, 2021, to August 23, 2021. The height and color of each bar represent the log frequency and ACE-S BFE change induced by a given RBD mutation.   {\bf d} Illustration of top 25 most observed RBD co-mutations. Here, the length of each bar represents the total ACE2-S BFE changes induced by a specific RBD co-mutation, the color of each bar represents the natural log frequency of each co-mutation, and the number at the side of each bar is the AI-predicted antibody disruption count.}
	\label{fig:barplot comut}
\end{figure}

The vaccine-resistant mutations are usually found along with other RBD mutations. Therefore, analyzing the time evolution of RBD co-mutations offers a better understanding of the mechanisms of vaccine-resistant mutations. Figures \ref{fig:barplot comut} {\bf a},  {\bf b}, and  {\bf c} illustrate the time evolution of 2, 3, and 4 RBD co-mutations with their corresponding BFE changes every 30 days. Here, the height and color of each bar represent the log10 frequency and total BFE change induced by a given RBD co-mutation. Considering the number of co-mutations is quite low in the year 2020,  the time range of analysis is set to [01/25/2021, 08/23/2021] for the time evolution analysis of 2 co-mutations. For 3 and 4 co-mutations, their time ranges are set to [02/04/2021, 08/23/2021] and [04/25/2021, 08/23/2021], respectively. In \autoref{fig:barplot comut} {\bf a},  red star marks the 2 co-mutations with significantly negative  BFE changes.  
 At the end of March 2021, vaccine-resistant mutation Y449D showed up with mutation N501Y in some genome isolates, resulting in a negative BFE change (-0.473kcal/mol) and a high antibody disruption count (98) for RBD 2 co-mutation [Y449D, N501Y]. However, its global frequency is relatively low. Since late April 2021, vaccine-resistant mutation Y449S showed up with N501Y, making RBD co-mutation [Y449S, N501Y]  one of the most prevailing vaccine-resistant co-mutations. \autoref{fig:barplot comut} {\bf d} shows the top 25 most observed RBD co-mutations, the length and color of each bar represent the total BFE change and the natural log of frequency of an RBD co-mutation. The number at the side of each bar is the count of antibody disruption. Among the 25 most observed RBD co-mutations, [Y449S, N501Y] is the only co-mutation with a significantly negative BFE change and extremely high antibody disruption count (94). Observing the evolution trajectory of [Y449S, N501Y] shows that the infectivity transmission pathway regulated by natural selection in the population level is the major evolution-driven force of SARS-CoV-2 mutagenesis before March 2021. Starting in January 2021, several vaccines were authorized for emergent use. Two months later, since many people have been protected by the vaccines, the mutations that disrupt the binding between the S and antibodies are able to transmit among vaccinated people, especially in countries with high vaccination rates. Such a vaccine-resistant pathway reduces the efficacy of vaccines and antibody therapies, indicating the combat with COVID-19 will be a prolonged battle. 

Similar  time evolution trajectories are drown for RBD 3 and 4 co-mutations (see Figures \ref{fig:barplot comut} {\bf b} and {\bf c}). 
Apparently, there are no vaccine-resistant 3 and 4 co-mutations at present, which indicates vaccine-resistant co-mutation [Y449S, N501Y] is quite unique.

\subsection{Vaccine-resistant mutations and vaccination rates in 12 countries}
\begin{figure}[ht!]
	\centering
	\includegraphics[width = 0.7\textwidth]{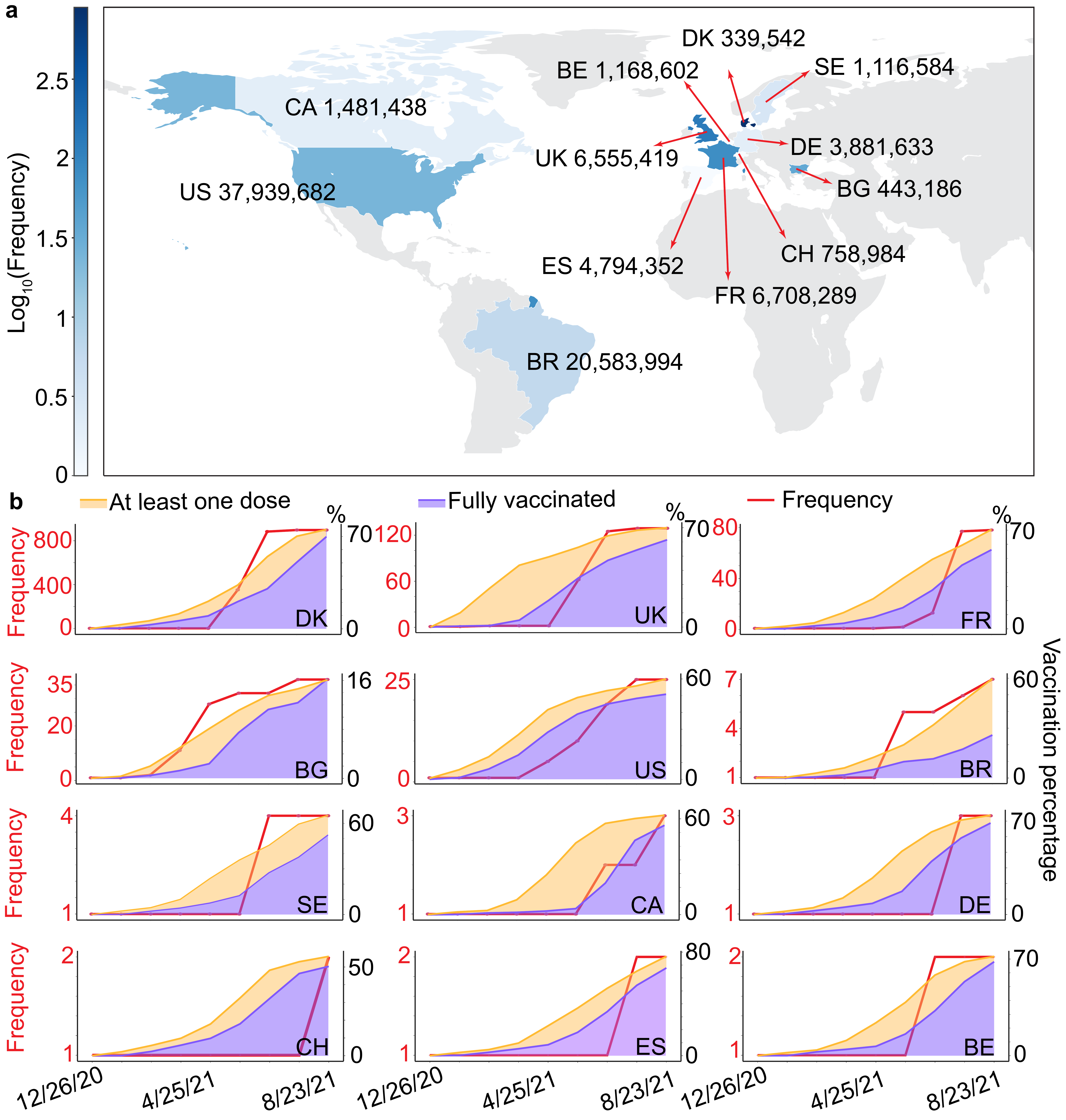}
	\caption{ {\bf a} Distribution of vaccine-resistant mutation Y449S. The color bar represents the log10 frequency of Y449S in 12 countries: Denmark (DK), the United Kingdom (UK), France (FR), Bulgaria (BG), the United States (US), Brazil (BR), Sweden(SE), Canada (CA), Germany (DE), Switzerland (CH), Spain (ES), and Belgium (BE). The number located at the side of the country shows the total positive SARS-CoV-2 cases up to August 31.  {\bf b} Time evolution of vaccination rate and the frequency of Y449S in 12 countries from December 26, 2020, to August 23, 2021. The data is collected per 30-day. The red line shows the frequency of mutation Y449S. The orange and purple areas represent at least one dose rate and fully vaccinated rate in each country. }
	\label{fig:lineplot}
\end{figure}

Analysis of the vaccination trends and vaccine-resistant mutations leads to a fundamental understanding of the transmission and evolution of vaccine-resistant mutations. We investigate the distribution and time evolution of vaccine-resistant RBD mutation Y449 in 12 countries. As the most observed vaccine-resistant RBD mutation, Y449S has been detected in 12 countries, including Denmark (DK), the United Kingdom (UK), France (FR), Bulgaria (BG), the United States (US), Brazil (BR), Sweden(SE), Canada (CA), Germany (DE), Switzerland (CH), Spain (ES), and Belgium (BE), as illustrated in  \autoref{fig:lineplot} {\bf a}. Here, 12 countries that Y449S was found in are in blue. The darker the blue is, the higher frequency of Y449S will be. The number on the side of each country is the total positive cases up to August 31, 2021. Although DK has the smallest positive cases among 12 countries, the frequency of Y449S is the highest. More than 800 patients carry vaccine-resistant mutation Y449S in DK. All of the Y449S-related cases are found in Europe and America, where the vaccination rates in those areas are relatively high. \autoref{fig:lineplot} {\bf b} shows the time evolution of vaccination ratio and the frequency of Y449S in the 12 countries as mentioned above in 30-day periods. The $x$-axis records the date, which ranges from 12/26/2020 to 08/23/2021. The left-hand side $y$-axis shows the frequency of Y499S (red lines), and the right-hand side $y$-axis shows the vaccination ratio. In addition, the orange region shows at least one dose ratio, while the purple region means the fully vaccinated ratio. It can be seen that Y449S was first found in BG and the US in December 2020. However, the frequency of Y449S in BG and the US is quite low before April 2021. After April 2021, Y449S has been quickly spread out to other ten countries. Among them, the total number of cases related to Y449S has a rapid increment tendency, especially in DK, the UK, and FR. Notably, all these three countries have relatively high vaccination ratios (over 70\% up to late August 2021). It is worthy to mention that the frequency of Y449S is low in DE, CH, ES, and BE, etc., which is mainly due to the first Y449-related case in these countries was detected after June 2021. Since then, Delta variants dominated in the prevailing variants, which gave Y449S a limited chance to spread out rapidly. Moreover, from \autoref{fig:lineplot} {\bf}, it can be seen that the frequency of Y449S has a similar growing tendency as the fully vaccinated ratio, suggesting that the vaccine-resistant mutations will gradually become one of the main evolution driven forces of SARS-CoV-2, especially in those areas with high vaccination rates. 

\section{Conclusion}
Due to the appearance of multiple mutations known to reduce the efficacy of antibody neutralization  generated by vaccines, it is vital to better comprehend the mechanisms of SARS-CoV-2 mutagenesis, which will be of paramount importance to understanding the transmission and evolution of SARS-CoV-2. The driven forces of mutagenesis can be categorized into three groups: 1) molecular-scale mechanism, 2) organism-scale mechanism, and 3) population-level mechanism. As an initial driven source of mutagenesis, the genetic information is changed by random shifts, viral proofreading, translation errors, etc., which all belong to molecular-scale mechanisms. Also, regulated by the host immune system, host gene editing, and rarely occurring host-viral recombination are two organism-scale mechanisms. The molecular- and organism-scale mechanisms provide a large number of candidate mutations in the SARS-CoV-2 genome, while it is the population-scale mechanism that determines what mutations become dominating.

Natural selection is a population-scale mechanism, which promotes the surge of the emerging SARS-CoV-2 variants by two complementary pathways: infectivity and vaccine resistance. The early stage of SARS-CoV-2 evolution was entirely dominated by infectivity-strengthening mutations. However, since late March 2021, once vaccines had provided protection to highly vaccinated populations, several vaccine-resistant mutations such as Y449S and Y449H have been observed relatively frequently. Considering there is still a good portion of the population who are not vaccinated, infectivity-strengthen mutations still dominate in the prevailing and future variants. However, antibody-resistant mutations will become a major mechanism of transmission once most of the populations carrying antibodies either through vaccination and infection. Our studies are valuable to the development of the next-generation vaccines and mAbs, which are of great importance in the long-term combat with SARS-CoV-2.

\section*{Data and model availability}
The SARS-CoV-2 SNP data in the world is available at \href{https://users.math.msu.edu/users/weig/SARS-CoV-2_Mutation_Tracker.html}{Mutation Tracker}. The most observed SARS-CoV-2 RBD mutations are available at \href{https://weilab.math.msu.edu/MutationAnalyzer/}{Mutaton Analyzer}. The TopNetTree model is available at \href{https://github.com/WeilabMSU/TopNetmAb}{TopNetmAb}. The detailed methods can be found in the Supporting Information {\color{teal} S1} and {\color{teal} S2}. The validation of our predictions with experimental data can be located in Supporting Information {\color{teal} S3}. The information of 130 antibodies with their corresponding PDB IDs, the SARS-CoV-2 S protein RBD SNP and non-degenerate co-mutations data can be found in Section {\color{teal} S5} of the Supporting Information. 

\section*{Supporting information}
The supporting information is available for 

\begin{enumerate}
    \item[S1] Supplementary data pre-processing and feature generation methods
    \item[S2] Supplementary machine learning methods
    \item[S3] Supplementary validation: validations of our machine learning predictions with experimental data.
    \item[S4] Supplementary table: the top 25 most observed S protein RBD mutations up to September 20, 2021. 
    \item[S5] Supplementary data: The Supplementary\_Data.zip contains four files: {\color{teal} S5.0.1}: antibodies\_disruptmutation.csv shows the name of antibodies disrupted by mutations. {\color{teal} S5.0.2}: antibodies.csv lists the PDB IDs for all of the 130 SARS-CoV-2 antibodies. {\color{teal} S5.0.3}: RBD\_comutation\_residue\_09202021.csv lists all of the SNPs of RBD co-mutations up to September 20, 2021. {\color{teal} S5.0.4}: Track\_Comutation\_09202021.xlsx preserves all of the non-degenerate RBD co-mutations with their frequencies, antibody disruption counts, total BFE changes, and the first detection dates and countries.
\end{enumerate}

\section*{Acknowledgment}
This work was supported in part by NIH grant  GM126189, NSF grants DMS-2052983,  DMS-1761320, and IIS-1900473,  NASA grant 80NSSC21M0023,  Michigan Economic Development Corporation, MSU Foundation,  Bristol-Myers Squibb 65109, and Pfizer.

\bibliographystyle{unsrt}
\bibliography{refs}
\end{document}